\documentclass[RNAAS]{aastex62}

\usepackage{amsmath} 
\newcommand{\angstrom}{\text{\normalfont\AA}}

\graphicspath{{./}{figures/}}

\begin{document}

\title{Wolf 1465: Not a Bright Dwarf Carbon Star}

\correspondingauthor{Bruce Margon}
\email{margon@ucsc.edu}

\author{Bruce Margon}
\affil{Department of Astronomy and Astrophysics, University of California, Santa Cruz, CA 95064, USA}

\author{Georgios Dimitriadis}
\affiliation{Department of Astronomy and Astrophysics, University of California, Santa Cruz, CA 95064, USA}

\keywords{stars: carbon -- proper motions}

\section{} 

Dwarf carbon (dC) stars - objects with prominent C$_{2}$ bands in their spectra, but luminosity near the main sequence – pose a variety of interesting problems. The long-standing presumption is that the carbon must have been transferred from a now invisible companion \citep{Green13ApJ...765...12G}, but empirical evidence to this effect has been elusive. Data that many, most or possibly all dC stars are binary, a concept suspected but unproven for decades, is finally starting to emerge from recent photometric \citep{Margon18ApJ...856L...2M}, spectroscopic \citep{Whitehouse18MNRAS} and astrometric \citep[][hereafter ``H18'']{Harris18AJ} observations. A remaining important issue, however, is metallicity. As yet only one dC star has a detailed abundance analysis, the prototype dC, G77-61, and it is extraordinarily metal poor, $[{\rm Fe/H}] = -4$ \citep{Plez05A&A...434.1117P}. Clearly a far larger sample is needed to probe this interesting abundance result, but even the brightest dC’s ($V\sim$14) require the largest telescopes for these analyses. Thus discovery of a previously unnoted very bright dC would be quite important.

Recently H18 have noted that $(I-J)$, $(J-K_{s}$) colors provide excellent segregation of dC stars from normal late-type dwarfs, and the 2MASS colors for the high proper motion star Wolf 1465 ( = LHS 466, = LFT 1419, = LTT 7381, =~G 155-27) fall in the dC regime. They point out that at $V=13.7$ \citep{Harrington80AJ.....85..454H}, this would make Wolf~1465 the brightest known dC, exceeding even the prototype G77-61. The recent {\it Gaia} DR2 parallax of 17 mas might also make it the closest such star. However, H18 note an important caveat: the UKIDSS colors disagree with 2MASS by 0.2 mag, indicating one of the measures is likely in error. The M$_{V}$ = 9.85 derived from the parallax is compatible with either a late type dwarf or a dC, so sheds no further light on the mystery.

Although Wolf 1465 first appeared in the literature almost a century ago \citep{Wolf25AN....223..231W}, and there are multiple astrometric measurements published, there is little published spectral data. \citet{Rodgers74PASP...86..742R} obtained but did not display a photographic spectrum, with the notation ``slightly weak lined'', which is typed as K4 by \citet{Augensen78Ap&SS..59...35A}. \citet{Bidelman85ApJS...59..197B} cites an unpublished photographic spectrum by G. P. Kuiper obtained years earlier, assigning spectral type K5. We are aware of no publication of an actual spectrogram, nor any reported observation since the use of solid state detectors became common.

To gain a definitive modern classification of Wolf 1465, we obtained the spectrum of the star on 2018 May 8 UT, using the Kast spectrograph of the Lick Observatory Shane telescope. The resulting 800 s exposure is shown in Figure 1; the spectral resolution is about 2.5 $\angstrom$. Although the modest resolution is not ideal for precise classification purposes, it is clear that Wolf 1465 is a normal mid-K star, in agreement with the unpublished photographic spectrograms of decades earlier. Prominent Ca lines (IR triplet, ${\lambda\lambda}4226, 6162$, H\&K), NaD, and the Mg~b band are present, but negligible Balmer or TiO absorption, bounding the classification on the warm and cool ends, respectively. There is certainly no sign of C$_{2}$ features. The M$_{V}$ = 9.85 derived from the parallax supports the interpretation as a dK star. It seems likely that H18\textsc{\char13}s suggestion that the 2MASS photometry is flawed is correct. Given the high proper motion of the star, 1260 mas yr$^{-1}$, it may even be possible that the object was misidentified there by an automated algorithm.

\begin{figure}[!ht]
\begin{center}
\includegraphics[width=0.95\textwidth]{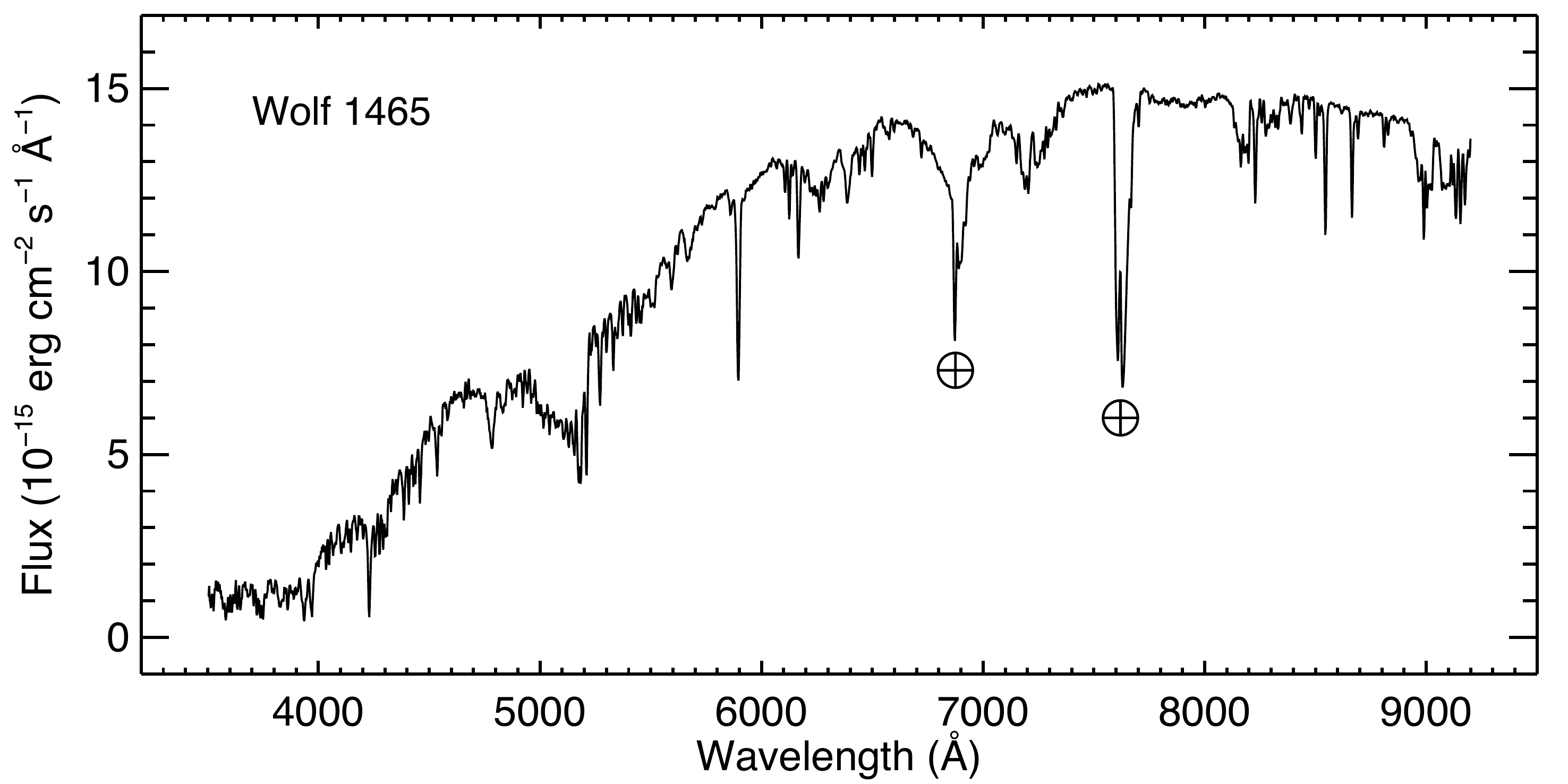}
\caption{The spectrum of Wolf 1465. The prominent spectral features, compatible with mid-K classification, are noted in the text; telluric bands are also marked. Flux calibration was obtained via observations of spectrophotometric standard stars, but should be regarded as approximate only, due to uncertain light losses at the slit.\label{fig:1}}
\end{center}
\end{figure}

Unfortunately Wolf 1465 is not the brightest dC star, and the original prototype from four decades past, G77-61 \citep{Dahn77ApJ...216..757D}, continues to hold this distinction.

\acknowledgments

The UCSC group is supported in part by NSF grant AST-1518052, the Gordon \& Betty Moore Foundation, the Heising-Simons Foundation, and by fellowships from the Alfred P. Sloan Foundation and the David and Lucile Packard Foundation to R. J. Foley. This work has made use of data from the European Space Agency (ESA) mission {\it Gaia} (\url{https://www.cosmos.esa.int/gaia}), processed by the {\it Gaia} Data Processing and Analysis Consortium (DPAC, \url{https://www.cosmos.esa.int/web/gaia/dpac/consortium}). Funding for the DPAC has been provided by national institutions, in particular the institutions participating in the {\it Gaia} Multilateral Agreement.

\bibliographystyle{aasjournal} 
\bibliography{rnaas}

\end{document}